# Emergence simulation of biological cell-like shapes satisfying the conditions of life using a lattice-type multiset chemical model

Takeshi Ishida [1*]

[1] Department of Ocean Mechanical Engineering, National Fisheries University; ishida08@ecoinfo.jp
* Correspondence: ishida08@ecoinfo.jp

**Simple Summary:** One of the great challenges in science is determining when, where, why, and how life first arose as well as the form taken by this life. In the present study, life was assumed to be (1) bounded, (2) replicating, (3) able to inherit information, and (4) able to metabolize energy. The various existing hypotheses provide little explanation of how these four conditions for life were established. Indeed, "how" a chemical process that simultaneously satisfies all four conditions emerged after the materials for life were in place is not always clear. In this study, a multiset chemical lattice model, which allows virtual molecules of multiple types to be placed in each cell on a two-dimensional space, was considered. Using only the processes of molecular diffusion, reaction, and polymerization and modeling the chemical reactions of 15 types of molecules and 2 types of polymerized molecules and using the morphogenesis rule of the Turing model, the process of emergence of a cell-like form with all three conditions except evolution ability was modeled and demonstrated. Thus, in future research, this model will allow us to revisit and refine each of the hypotheses for the emergence of life.

**Abstract:** Although numerous reports using methods such as molecular dynamics, cellular automata, and artificial chemistry have clarified the process connecting non-life and life on protocell simulations, none of the models could simultaneously explain the emergence of cell shape, continuous self-replication, and replication control solely from molecular reactions and diffusion. Herein, we developed a model to generate all three conditions, except evolution ability, from hypothetical chains of chemical and molecular polymerization reactions. The present model considers a 2D lattice cell space, where virtual molecules are placed in each cell, and molecular reactions in each cell are based on a multiset rewriting rule, indicating stochastic transition of molecular species. The reaction paths of virtual molecules were implemented by replacing the rules of cellular automata that generate Turing patterns with molecular reactions. The emergence of a cell-like form with all three conditions except evolution ability was modeled and demonstrated using only molecular diffusion, reaction, and polymerization for modeling the chemical reactions of 15 types of molecules and two types of polymerized molecules. Furthermore, controlling self-replication is possible by changing the initial arrangement of a specific molecule. In summary, the present model is capable of investigating and refining existing hypotheses on the emergence of life.

## 1. Introduction

In science, one of the great challenges is clarifying when, where, why, and how life first arose as well as the form that the first life took. To this end, research related to the origin of life has long been conducted, and a vast number of papers and books, e.g. [1,2], have been published. "When" life came into being is a question that has been studied from two perspectives: (a) from a geological perspective, i.e., examining when the Earth's environment became compatible with the survival of life and (b) from the perspective of studying the record of life using fossil sources. According to the geology of Earth, various hypotheses have been postulated, e.g., life arose 4.4 billion years ago when the environment and conditions were ripe for its emergence. When attempting to determine the emergence of first life from fossils, traces of life were found to exist 3.8 billion years ago; however, notable limitations exist to tracing fossils to ascertain the emergence of life. Several hypotheses have been postulated to explain "where" the first life arose, which may have included hydrothermal vents, the submarine subsurface, and land-based hot springs and geysers, which each demonstrate various advantages and disadvantages. For each hypothesis, many studies have been conducted to examine the materials and physical conditions necessary for the composition of life in the proposed location. As for "why" life emerged, no one hypothesis directly answers the question, but research on nonequilibrium thermodynamics, such as dissipative structure and entropy studies, may help find an answer. The concept of life as a physical phenomenon began with Schrödinger, who proposed "negative entropy" [3], and with the dissipative structure proposed by Prigogine [4]. Considering life as dissipative structure phenomenon, approaching the "why" question of the origin of life is possible by exploring the conditions in which dissipative structures emerge. Further, this question might be answered by studying complex adaptive systems or self-organizing phenomena.

Concerning the questions of "what" was the first life and "how" did it come into being, several hypotheses have been proposed, including the RNA world hypothesis [5]. Indeed, two major classifications exist: the "RNA world hypothesis," suggesting that RNA with an autocatalytic function was the first life, and the "metabolic world (chemical evolution) hypothesis" [6], which points to the catalytic function of mineral surfaces etc. Other hypotheses, such as the "protein world hypothesis" [7], have been suggested, and research is being conducted to predict the first forms of life from the oldest living organisms. An example of life emergence scenario would be a volcanic hot spring, where life-supporting organic matter is concentrated and polymerized in a dry–wet cycle while simultaneously being trapped in fatty acid compartment [8]. However, the process by which this spherical fat retains the ability to metabolize and replicate is not yet clear. The same can be said for the hydrothermal vent hypothesis [9]. Even if the first life was self-replicating RNA, the process of RNA acquiring membranes and evolving to metabolize and self-replicate has yet to be clarified. Moreover, the other hypotheses provide little explanation of how the four conditions for life were established. In the various hypotheses, the emergence of these conditions is vaguely explained as "the result of a long process of trial and error" or the function of life arising when a "certain level of complexity" has been reached. Therefore, it is not always clear "how" a chemical process that simultaneously satisfies all conditions emerged after the materials for life were in place.

When considering the origin of life, considering the conditions that might determine the emergence of life is necessary. Accordingly, the definition of life must first be clarified. However, this definition is also widely debated. A major discussion or literature review has been avoided here; instead, I considered the four conditions of life that are generally accepted, i.e., that life is (1) bounded, (2) replicating, (3) able to inherit information, and (4) able to metabolize energy. Any one of these conditions does not define life; to reach the level of life, each of these conditions must be fulfilled simultaneously. Life cannot be synthesized spontaneously by simply accumulating a high concentration of the materials necessary for life.

Many studies attempted to clarify this process from non-life to life, beginning with Opalin's coacervate theory [10] and subsequently through Miller's experiment [11], which led to numerous studies investigating the process of creating the materials of life. Intermediates leading from non-life to life are sometimes called protocells. Areas of research building protocells include: (a) approaches to developing a protocol in an actual experimental system, and (b) computational scientific approaches. With respect to (a), attempts to create the protocell with the essential functions of a cell are underway [12].

For example, In the field of synthetic biology, research is being undertaken to synthesize life from genetic information. The Craig Bender Institute reported that synthetic life was successfully created by synthesizing a bacterial DNA molecule and transferring it to another cell [13]. Another recent study reported the experimental construction of a protocell [14].

In contrast, computational approaches (b) include: Calculating protocells from (b-1) molecular dynamics models, (b-2) based on probabilistic models or discrete models such as cellular automata and artificial chemistry. The research field of "artificial life," which seeks to understand the essential nature of life based on computer simulations, flourished since the 1990s. In this field, research on models that behave like life has been conducted from a mathematical perspective. Regarding (b-1), molecular dynamics (MD) is a modeling method for calculating molecule behavior that requires high computation to calculate the behavior of a small number of molecules, as it solves each individual molecule mechanically. This method can calculate only a limited time within a limited space, making it difficult to compute a single whole living cell. For this reason, reducing the number of degrees of freedom is necessary to reduce computational complexity. There are many reports on coarse-grained models of MD, including computational examples of the self-organizing structure of cell membranes [15,16], of cell membrane division [17], and simulations of nanoscale mechanisms for nanovesicles' budding and fission [18]. In addition, stochastic-deterministic simulations over a cell cycle using a whole-cell fully dynamical kinetic model have been reported [19]. These studies are simulations for each protocell construction process, but do not refer to the emergence of protocells from molecules and acquisition of the four conditions of life.

Regarding (b-2), cellular automaton (CA) model is a discretization model of time and space in which the state of each lattice cell is determined by the state of its neighbors. A well-known example of a CA is Conway's Life Game [20]. In addition, a study examined the growth and motility of protocells using a CA model [21]. In another study using a virtual particle reaction field with specific particle transformations and temperature fields, Ishida [22] simulated the process of cell shape emergence and partial replication through a combination of particle transformations and binding. However, even with this model, results allowing for stable cell shape and continuous self-replication could not be obtained. Almost all CA models exhibit one state per cell (despite the few types of CA with multiple fractional states) and have difficulty handling multiple types of molecules, making it challenging to handle complex chemical reactions and molecular structures.

Regarding research on artificial chemical models in (b-2), the computational representation of chemical reactions of molecules has been modeled as a multiset chemical model known as the "artificial chemical model" [23]. Multiset is a concept of a mathematical class plus multiplicity. Using the multiset concept, constructing a model that considers both the number and type of molecules is possible. For example, considering molecule *a* and molecule *b*, a multiset representation of the state with three of molecule *a* and two of molecule *b* would be {*a, a, a, a, b, b*}. When molecule *a* and molecule *b* transform into molecule *c* through chemical reactions, this can be expressed via the following multiset rewriting rule: {*a, a, a, b, b*} → {*a, c, c*}. Regarding research on protocells by artificial chemistry, Kruszewski et al. [24] validated self-reproducing metabolisms with a minimalistic artificial chemistry based on a Turing-complete rewriting system called combinatory logic. Fellermann [25] performed a dissipative particle dynamics simulation, combining self-assembly processes with chemical reaction networks. Moreover, Hutton [26] constructed a replication model by the interaction of two-dimensional particle swarms.

From a macro perspective, it is common to express the number of molecules in terms of concentration, which is a continuous quantity, and to express changes in concentration over time using differential equations. However, in many cases, such as in cells, a relatively small number of molecules or their polymers express important functions, and limitations exist to using concentration alone. There is also research on protocells by system analysis [27,28].

It is also necessary to look at the protocol from the perspective of emergence. Many mathematical studies have focused on the emergence of various life forms from local interactions in reaction–diffusion equations, including the Turing pattern model [29]. For example, Young's model [30] is a two-dimensional totalistic model that bridges reaction–diffusion equations and the CA model, producing Turing patterns. Adamatzky et al. [31] studied a binary-cell-state eight-cell neighborhood two-dimensional CA model with semitotalistic transitions rules, and Dormann et al. [32] used a two-dimensional outer-totalistic model with three states to produce a Turing-like pattern, whereas Tsai et al. [33] analyzed a self-replicating mechanism in Turing patterns with a minimal autocatalytic monomer–dimer system. By improving the external sum rule type CA, in addition to the concept of the rules of the Game of Life, Ishida [34] showed that various shapes such as self-replication, growth, move, and branching, which does not arise from a pure Turing pattern model, can emerge and be controlled using a small number of parameters. Further, Ishida [35] showed, using computer modeling that Turing patterns can be formed with time-course changes and adjacent interactions of only one type of virtual molecule. However, although these models can simulate the emergence of relatively simple shapes, such as spots and stripes, they cannot reproduce the emergence process of complex structural systems, such as biological cells that retain genetic information and self-replicating functions.

As explained above, none of the models could simultaneously achieve the emergence of cell shape, continuous self-replication, and replication control solely from molecular reactions and diffusion. In the present study, to overcome the shortcomings of the conventional CA model, which cannot represent complex constructs such as proteins, a model that allows virtual molecules of multiple types to be placed in each cell on a 2D space was considered (Figure 1). This model is a lattice-type artificial chemical model, capable of describing a wide variety of states and interactions in a limited number of lattice cell spaces, including the migration (diffusion), transformation (chemical reaction), and linkage (polymerization) of virtual molecules. In addition, the parameters required for the model can be described using the number or ratios of molecules and can be internalized within the calculation field. The current study shows that emerging a phenomenon that satisfies the three conditions except evolution ability is possible: the emergence, maintenance, and self-replication of a cell-like boundary with the ability to retain genetic information. However, the ability of evolution is an issue for future research.

Although 15 types of artificial molecules are assumed in the model presented here, the number of chemical reaction combinations among these molecule types is enormous; thus, identifying the reactions among these combinations that are responsible for the life emergence phenomenon is difficult. Therefore, the algorithm used for generating Turing patterns identified in Ishida's model [34,35], was modified to procedures for diffusion and reaction of 15 different virtual molecules and their polymerization. Using only the processes of molecular diffusion, reaction, and polymerization, this model was able to realize a process in which metabolism begins, boundaries are created, these boundaries replicate, and the information necessary for the maintenance of form (equivalent to genetic information) is retained.

This study neither provides a new hypothesis of origin of life nor verifies a conventional hypothesis. It does, however, propose a simple yet concrete mathematical model to explain the point that, until now, has only been vaguely expressed as "the conditions for life were established as a result of a long process of trial and error."

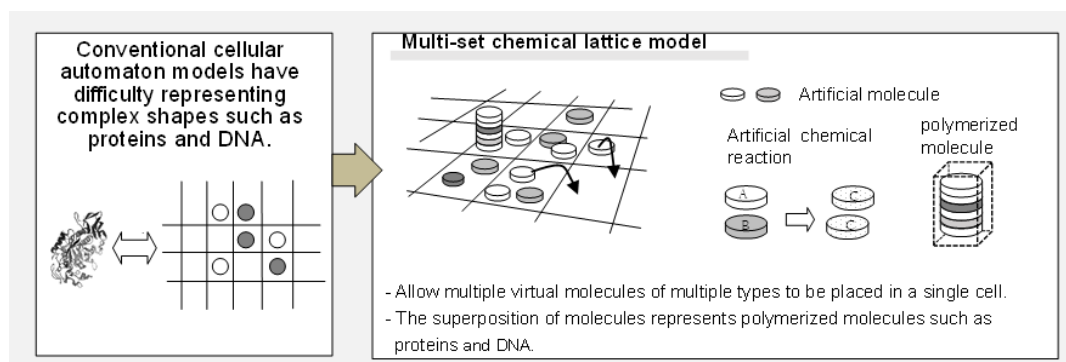

**Figure 1.** Schematic overview of the multiset chemical lattice model. This model allows multiple virtual molecules of multiple types to be placed in a single cell on a two-dimensional lattice space and is capable of describing a wide variety of states and interactions in a limited number of cells, e.g., diffusion, chemical reactions, and polymerization of molecules.

## 2. Materials and Methods

### 2.1. Model Configuration

In this study, a model was constructed that can generate the three conditions of life from hypothetical chains of chemical reactions and molecular polymerization reactions. In the present model, a 2D lattice cell space is considered, virtual molecules are placed in each cell, and the reaction of the molecules in each cell is based on the concept of a multiset rewriting rule in which the transition of molecular species occurs stochastically. Three-dimensional models will be the subject of future research. An overview of the model is shown in Figure 2. In each cell, the number of molecules is recorded for each molecular species.

The method for modeling the diffusion and reaction processes of each molecule in the lattice cell will now be described. As shown in Figure 2A, molecular diffusion can be represented by the exchange of molecular numbers between adjacent cells. In the current model, the diffusion of molecules and polymerized molecules in each cell is represented by the process shown in Figure 2A. Briefly, the process by which molecules are replaced from a cell with a large number of molecules to a neighboring cell with a small number of molecules is termed molecular diffusion. The residual rate was defined here as an alternative parameter to the diffusion coefficient.

The residual ratio $r$, a parameter of each molecular type, is the fraction of unmoved molecules in each cell. This parameter is a molecule-specific attribute value, fixed for all lattice cells and all-time steps. As shown in Figure 2A, the following calculations are performed, wherein $b_{n,t}$ is the number of molecules in cell $n$ at time $t$: (1) a proportion, $b_{n,t} \times (1 − r) \times 1/6$, of the molecules diffuse toward the six adjacent cells evenly; (2) the residual, $b_{n,t} \times r$, molecules remain in the original cell. If the number of molecules is not an integer multiple of six, the remainder of the molecules are distributed between adjacent cells with equal probability.

As shown in Figure 2B, chemical reactions can be represented by rewriting the number of molecular types based on reaction probabilities, based on the multiset rewriting rule. In the present model, 15 types of molecules were assumed to replace the Ishida's algorithm [34] with molecular reactions, as described below. Because this is an artificial chemical model, it does not correspond to real molecules, and molecular species are represented as "molecule 1," "molecule 2," etc. The chemical reaction is assumed to occur at each time step with a certain probability of change in the number of molecules in each cell, as shown in Figure 2B. For example, in the case of a reaction involving molecule $A$ and molecule $B$, molecule $A$ is converted to molecule $B$ with the reaction probability $p$.

Assuming that the number of molecule *A* at time **t** is $A_n$ and the number of molecule *B* is $B_n$, the number of molecules at the next time step is as follows:

$$A_{n,t+1} = A_{n,t} \times (1-p)$$
$$B_{n,t+1} = B_{n,t} + A_{n,t} \times p$$

Similarly, the reaction *molecule A + molecule B → molecule C + molecule D* is conducted with an increase or decrease in the number of molecules, according to the reaction probability. Accordingly, chemical reactions describe the change in the number of molecules of each molecular species.

The 15 molecule types set up in the present model demonstrate the following roles.

- Molecule 1: the material to be converted into each molecule (initially, a large number of such molecules are placed in the lattice space).
- Molecules 2 and 3: correspond to diffusing substances in the Turing pattern model (the difference of diffusion coefficients is expressed by the difference in their residual rate).
- Molecules 4 and 5: substances that change from molecules 2 and 3 during diffusion, respectively.
- Molecules 6 and 7: materials that constitute "polymerized molecule 1," which is a polymer of molecules 6 and 7 (the ratio of molecules 6 and 7 represents the morphology parameter *w*).
- Molecules 8, 9, 10, and 11: describe the transition equation of the Ishida model [34] [equation (1), described below] in chemical reactions.
- Molecule 12: the material that makes up "polymerized molecule 2," which represents the boundary of the cell.
- Molecules 13, 14, and 15: describe the transition equation of the Ishida model [34] [equation (2), described below] in chemical reactions.

A large number of molecule 1, i.e., the material to be converted into each type of molecule, is initially placed in the lattice cell space. Then, the distribution of molecule 1 in lattice space changing stochastically to other molecules is modeled. These chemical reactions are assumed to involve changes into several molecular species before changing back into molecule 1, creating a cyclic chemical reaction world. For example,

Molecule 1 → Molecule 2 → Molecule 3
Molecule 3 + Molecule 4 → Molecule 5 + Molecule 6
Molecule 6 → Molecule 1

Within the lattice space, the creation of new molecules is not assumed, and the total number of molecules combined with molecular types is assumed to be constant.

Macromolecules, such as cell membranes and genes, also play essential roles in the processes of life emergence; therefore, methods to represent molecular polymers are necessary. In the present model, a virtual box is filled with virtual molecules collectively, as shown in Figure 2C. The virtual empty box in the space awaits the polymerization of molecules when the box is filled with molecules. When the box is emptied after the removal of molecules, it represents a state in which the polymerized molecules are broken down and returned to small molecules. Furthermore, the box itself was modeled to be diffuse in the lattice space.

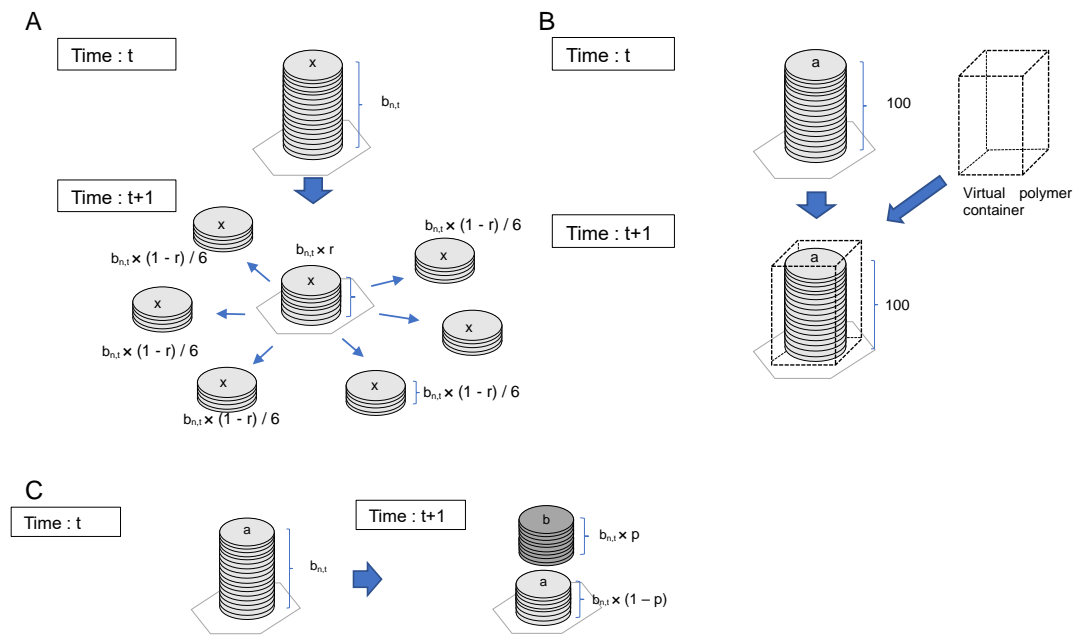

**Figure 2.** Schematic of molecule diffusion, chemical reactions, and the virtual box model representing the state of polymers. (A) Molecular diffusion was represented by the exchange of molecule numbers between adjacent cells. (B) Chemical reactions are represented by rewriting the number of molecule types based on reaction probabilities. One example of a reaction from molecule $a$ to molecule $b$ with reaction rate $p$ is shown. (C) A virtual box was assumed to be filled with virtual molecules collectively. Next, the empty box in the lattice space was placed to represent the polymerization of molecules when the box was filled with molecules. When the box was emptied after the removal of molecules, it represents a state in which the polymerized molecules are broken down and returned to small molecules.

*2.2. Conditions for Chemical Reactions to Guide Cellular Emergence*

Based on the foundation of the model described above, multiple types of molecules are placed in lattice cell space to calculate diffusion and chemical reaction processes. However, as the number of molecule types increases, the number of possible reaction path combinations increases enormously, and identifying the combination that will cause the desired reaction becomes difficult (i.e., the objective of creating a cell shape that meets the conditions of life).

In this study, we used two algorithms of Ishida [34] to facilitate the emergence of Turing patterns. Replacing these algorithms with equivalent molecular reactions, the molecular species and chemical reactions were set up. The algorithm for the formation of cellular shapes in Ishida's model [34] is based on the transformation of the differential equation of the Turing pattern into the transition rule of the CA model. The models of Markus et al. [36] and Young [30] are Turing pattern models with CA models. The Young model is a simple model for reproducing animal body patterns; Ishida's [34] is an applied Young model.

Young's model uses a real number function (Figure 3A), to derive distance effects and contains two constant values, $u_1$ (positive) and $u_2$ (negative). The calculation begins by randomly distributing black cells on a rectangular grid (Figure 3B). For each cell at position $R_0$ in the 2D field, the next cell state (black or white) of $R_0$ then results from the sum of the function values of black cells at $R_i$ positions. $R_i$ is assumed to be a black cell within a radius $r_2$ from the $R_0$ cell, and $i$ is the number of black cells within radius $r_2$ from $R_0$. The function $v(r)$ is a continuous step function, as shown in Figure 3A. The activation area, indicated by $u_1$, exhibits a radius of $r_1$, whereas the inhibition area, indicated by $u_2$, demonstrates a radius of $r_2$ ($r_2 > r_1$) (Figure 3B). At position $R_0$, when $R_i$ is assumed to be a grid within $r_2$, $v(|R_0 - R_i|)$ function values are summed. Function $|R_0 - R_i|$ indicates the distance between $R_0$ and $R_i$. If $\sum_i v(|R_0 - R_i|) > 0$, the grid cell at point $R_0$ is marked as a black cell. If $\sum_i v(|R_0 - R_i|) < 0$, the grid cell becomes a white cell. If $\sum_i v(|R_0 - R_i|) = 0$, the grid cell does not change state [30]. Young showed that a Turing pattern can be generated using these functions. Moreover, spot patterns or stripe patterns can be created with relative changes in $u_1$ and $u_2$.

Ishida [34] modified the model of Young, converting it into the following equivalent model in which Turing patterns also emerge. In Young's model, let $u_1 = 1$ and $u_2 = w$ (here $0 < w < 1$). If the state of the cell is set to *0* (white) or *1* (black), the model can be arranged as indicated below. The state of cell *i* is expressed as $c_i(t)$ ($c_i(t) = [0, 1]$) at time *t*. The following state $c_i(t + 1)$ at time $t + 1$ is determined by the states of the neighboring cells. Here $N_1$ is the sum of the states of the domain within the $s_1$ mesh of the focal cell. Similarly, $N_2$ is the sum of the states of the domain within the $s_2$ mesh of the focal cell, assuming that $s_1 < s_2$.

$$N_1 = \sum_{i=1}^{S_1} c_i(t)$$
$$N_2 = \sum_{i=1}^{S_2} c_i(t)$$

Here, $S_1$ and $S_2$ are the numbers of cells within the $s_1$ and $s_2$ meshes of focal cells, respectively. In addition, $s_2 = 2 \times s_2$ was assumed in the paper [34]. The next time-state of the focal cell is determined using expression (1). Here, two parameters, namely *w* and *s*, determine the Turing pattern:

$$\text{Cell state at the next time step} = \begin{cases} 1 : if \ N_1 > N_2 \times w \\ Unchanged : if \ N_1 = N_2 \times w \\ 0 : if \ N_1 < N_2 \times w \end{cases} \quad (1)$$

Furthermore, in Ishida's model [34], the parameters $w_1$ and $w_2$ are used to modify expression (1) to transition to state 1 only in a specific range. Thus, Ishida [34] shows that not only simple Turing patterns but also a variety of dynamic patterns (Conway's life game like patterns) are emergent.

$$\text{Cell state at the next time step} = \begin{cases} 0 : if \ N_1 > N_2 \times w_2 \\ 1 : if \ N_1 > N_2 \times w_1 \\ Unchanged : if \ N_1 = N_2 \times w_1 \\ 0 : if \ N_1 < N_2 \times w_1 \end{cases} \quad (2)$$

$$N_1 = \sum_{i=1}^{S} c_i(t)$$
$$N_2 = \sum_{i=1}^{T} c_i(t)$$

Here, $1 > w_2 > w_1 > 0$, and $N_1$ is the sum of the states of the domain within *s* meshes of the focal cell. Similarly, $N_2$ is the sum of the states of the domain within *t* meshes of the focal cell, assuming that $s < t$. *S* is the number of cells within *s* meshes from the focal cell, and *T* is the number of cells within *t* meshes from the focal cell. As in Young's model, "Unchange" was used as the equality in expression (2) to prevent all cells becoming state 1 or 0 in the next time step when $N_1 = 0$ and $N_2 = 0$.

Young's model and Ishida's modified model contain simple algorithms; however, to determine the transition of a particular cell, accumulating the information of distant cells within a certain range is necessary, such as $N_1$ and $N_2$. If the transition model depends on information from these distant cells, replacing the algorithm with a model of multiset chemical reactions is difficult. Therefore, a model is needed in which the transitions of a cell are determined solely by information within that cell and its neighbors.

In the second model proposed by Ishida [35], information from distant cells is not used, and the token changes in 16 steps as shown in Figure 3C, indicating that Turing patterns can be generated only by exchanging information with neighboring cells. However, the token's 16-step change model can be completed theoretically by assuming 16 chemical reactions, although the model becomes very complicated. Additionally, this Ishida model [35] determines the state of the space according to the number of tokens in each stage of each cell after the tokens diffused for a certain period of time. The diffusion process and transition decision process are separated in time, and the token placement is also reset; thus, the model is not a continuous time reaction model. In addition, a decision routine involving the shape parameter $w$ according to the number of tokens is conducted, and this routine must be expressed in chemical reactions.

In the present model, the process of diffusion is assumed to be the diffusion of molecules with two different diffusion coefficients in accordance with the Turing pattern model rather than the token change model. As described below, it was considered two types of molecules with different diffusion coefficients and used the process in which these molecules are transformed into another molecule during diffusion.

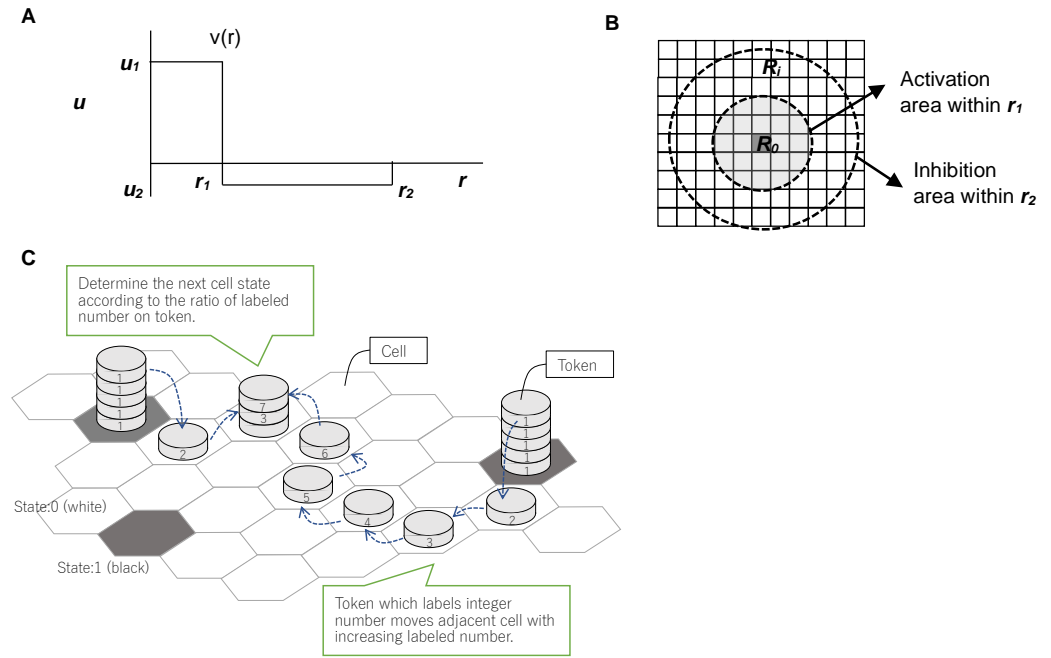

**Figure 3.** Outline of Young's model and Ishida's token model [35]. **(A)** In Young's model, function $v(r)$ is a continuous step function representing the activation area and inhibition area. **(B)** In Young's model, the activation area demonstrates a radius $r_1$, and the inhibition area demonstrates an outer radius $r_2$. **(C)** In a two-dimensional hexagonal grid, two-state cells (black and white) are assumed. The modification process of the tokens that are produced in the black cells is shown. The number labeling the token is modified up to X times in each cell, and some tokens remain in their cells without modification. The frequency distribution of token numbers determines the subsequent state of each cell. Here, X is the maximum number that can be attached to a token and is a positive integer.

The chemical reactions in the current model are described below. As an initial condition, 1,000,000 molecules of molecule 1, 50,000 molecules of molecule 6 and 7, respectively, and 100,000 molecules of molecule 12 were placed in each cell throughout the lattice space. In addition, 100 molecules of molecule 2 and 3, respectively, were placed in several cells in the center of lattice space as the initial arrangement. Molecules 2 and 3 correspond to the two types of diffusing substances in the Turing pattern model. Then, these two molecules diffuse into adjacent cells at each time step while simultaneously changing into molecules 4 and 5 at a constant reaction rate as follows:
(Reaction)
Molecule 2 → Molecule 4 (reaction rate: 5%)
Molecule 3 → Molecule 5 (reaction rate: 2%)

The reaction probability is set higher than that of molecule 3 → 5 to allow molecule 2 to react quickly at a short distance and change to molecule 4. The residual rates were set as molecule 2 = 0.75 and molecule 3 = 0.05, with molecule 2 diffusing slowly and molecule 3 spreading quickly. These rates result in higher concentrations of molecules 4 and 5 at short and long distances, respectively, reproducing the same situation as the two types of morphogens with different diffusion coefficients in the Turing pattern model.

Next, reactions were set up to produce as many of molecules 8, 9, and 11 as the number of molecules 4 and 5 in each cell. Molecules 8, 9, and 11 are used to determine the formation of polymerized molecule 2 through the reactions described below. In expression (1) of Ishida [34], molecule 4 (and 8) corresponds to $N_1$ in expression (1) and molecule 5 (and 9) corresponds to $N_2 - N_1$.
(Reaction)
Molecule 4 + Molecule 1 → Molecule 4 + Molecule 8 (reaction rate: 100%)
Molecule 4 + Molecule 1 → Molecule 4 + Molecule 11 (reaction rate: 100%)
Molecule 5 + Molecule 1 → Molecule 5 + Molecule 9 (reaction rate: 100%)

Molecules 8 and 9 react to molecule 10 with the ratio of molecule 6 in polymerized molecule 1. In expression (1), the number of molecule 10 corresponds to $N_2 \times w$.
(Reaction)
Molecule 8 → Molecule 10 (reacts at the ratio of molecule 6 in polymerized molecule 1)
Molecule 9 → Molecule 10 (reacts at the ratio of molecule 6 in polymerized molecule 1)

Here, molecules 8 and 9 are reset to molecule 1:
(Reaction)
Molecule 8 → Molecule 1 (reaction rate: 100%)
Molecule 9 → Molecule 1 (reaction rate: 100%)

The following reaction retains molecule 10 or 11, depending on which is more abundant, and the other molecule is reduced to molecule 1.
(Reaction)
Molecule 10 + Molecule 11 → Molecule 1 + Molecule 1 (reaction rate: 100%)

In each cell, the presence of molecule 11 is the same as the determination of the transition expression (1). The number of molecule 4 corresponds to the number of state 1 cells ($N_1$) in the expression (1). Molecule 10, which is reduced from molecules 8 and 9 with the ratio of molecule 6 in polymerized molecule 1, corresponds to $N_2 \times w$. The number of molecule 10 is compared with the number of molecule 11 (the same number as molecule 4). If molecule 11 exists in the lattice cell, the same condition of $N_1 > (N_2 \times w)$ is satisfied. To be equivalent to expression (1), a series of reactions must proceed quickly in a short period; thus, both reaction rates are set to 100%.

(Synthesis reaction of polymerized molecule 1)
If molecule 11 is present in the cell and an empty box is found, polymerized molecule 1 is synthesized in the same molecule 6/molecule 7 ratio as that of the other polymerized molecule 1 in the cell. Polymerized molecule 1 plays the role of an informant and holds information related to morphological parameters [ratio of molecule 6/molecule 7, corresponding to $w$ in expression (1)]. As polymerized molecule 1 is synthesized and diffuses into the lattice space, information is propagated outside the cellular domain. This represents the prototype of genetic information. The transmission method of genetic information that determines cell shape differs from that of real-life forms, in which genetic information is stored only within the biological cell, but it is more appropriate for early unstable cells to use informants filled within the surrounding space. Thus, the morphological parameter $w$ in the Ishida model [34] is embedded in the lattice space in the form of the ratio of molecules in polymerized molecule 1.

Furthermore, in the self-replication control model of Ishida [34], with modification of the transition formula as shown in expression (2), cell shape replication emerges, and various pattern formations can be controlled. To achieve the equivalent action of Ishida's model, the chemical reaction equivalent to the algorithm in expression (2) was set up as follows:

(Reaction)

Molecule 14 + Molecule 11 → Molecule 14 + Molecule 1 (reaction rate: 100%)
Molecule 15 + Molecule 11 → Molecule 15 + Molecule 13 (reaction rate: 100%)
Molecule 13 + Molecule 11 → Molecule 11 + Molecule 11 (reaction rate: 100%)

Molecules 14 and 15 were placed uniformly with three and nine molecules, respectively, in each cell. The above three reactions result in a process wherein molecule 13 remains when >4 molecules and <18 molecules (9 × 2) of molecule 11 exist. The algorithm corresponding to expression (2), in which the reaction proceeds when the number of molecules is within a certain range, can then be realized in molecular reactions. It was further assumed that when molecule 13 is present, polymerized molecule 2, which corresponds to the membrane molecule, is generated while, simultaneously, new molecules 2 and 3 are generated from molecule 1.

(Reaction)

Molecule 11 + Molecule 1 → Molecule 11 + Molecule 2 (reaction rate: 100%)
Molecule 11 + Molecule 1 → Molecule 11 + Molecule 3 (reaction rate: 100%)
Molecule 13 + Molecule 1 → Molecule 13 + Molecule 2 (reaction rate: 100%)
Molecule 13 + Molecule 1 → Molecule 13 + Molecule 3 (reaction rate; 100%)
Molecule 14 + Molecule 1 → Molecule 14 + Molecule 2 (reaction rate; 100%)
Molecule 14 + Molecule 1 → Molecule 14 + Molecule 3 (reaction rate: 100%)

(Synthesis reaction of polymerized molecule 2)

If 13 molecules exist in the cell with an empty box, 100 molecules of molecule 12 are linked together to synthesize polymerized molecule 2. Although in CA-based models, such as Ishida's [34], lattice cell states are color-coded to show 0 and 1, the present model shows the state of a lattice cell by the presence of polymerized molecule 2, which can represent cell shape via localized clustering of this polymerized molecule.

After the above reactions proceed in each cell, molecular removal reactions are assumed to proceed to establish cyclical reaction networks.

(Reaction)

Molecule 4 → Molecule 1 (reaction rate: 5%)
Molecule 5 → Molecule 1 (reaction rate: 5%)
Molecule 8 → Molecule 1 (reaction rate: 100%)
Molecule 9 → Molecule 1 (reaction rate: 100%)
Molecule 10 → Molecule 1 (reaction rate: 5%)
Molecule 11 → Molecule 1 (reaction rate: 5%)
Molecule 13 → Molecule 1 (reaction rate: 75%)

Further, it is natural for polymerized molecules to decompose at a certain rate. However, for simplicity, polymerization molecule 1 was not set to decompose in this study. In contrast, if molecule 13 is not present, polymerized molecule 2 will be degraded back to molecule 12 with a 5% probability.

By setting the molecular diffusion and reaction in each lattice as described above, it is possible to produce algorithms similar to Ishida's [34] and generate a variety of forms, including Turing patterns. Thus, the system sets up a cyclical reaction path wherein molecules decompose at a certain ratio and return to molecule 1, and this chain of reactions is thought to model metabolism at the same time. All reaction pathways are shown in Figure 4.

In contrast, the energy balance is assumed to be uniform for all reactions because of artificial molecule reactions. Additionally, the assumption exists that the lattice space is sufficiently energized for chemical reactions to proceed. Therefore, if the molecules necessary for the reaction are present in each cell, the reaction is assumed to proceed with the probability indicated by the reaction rate.

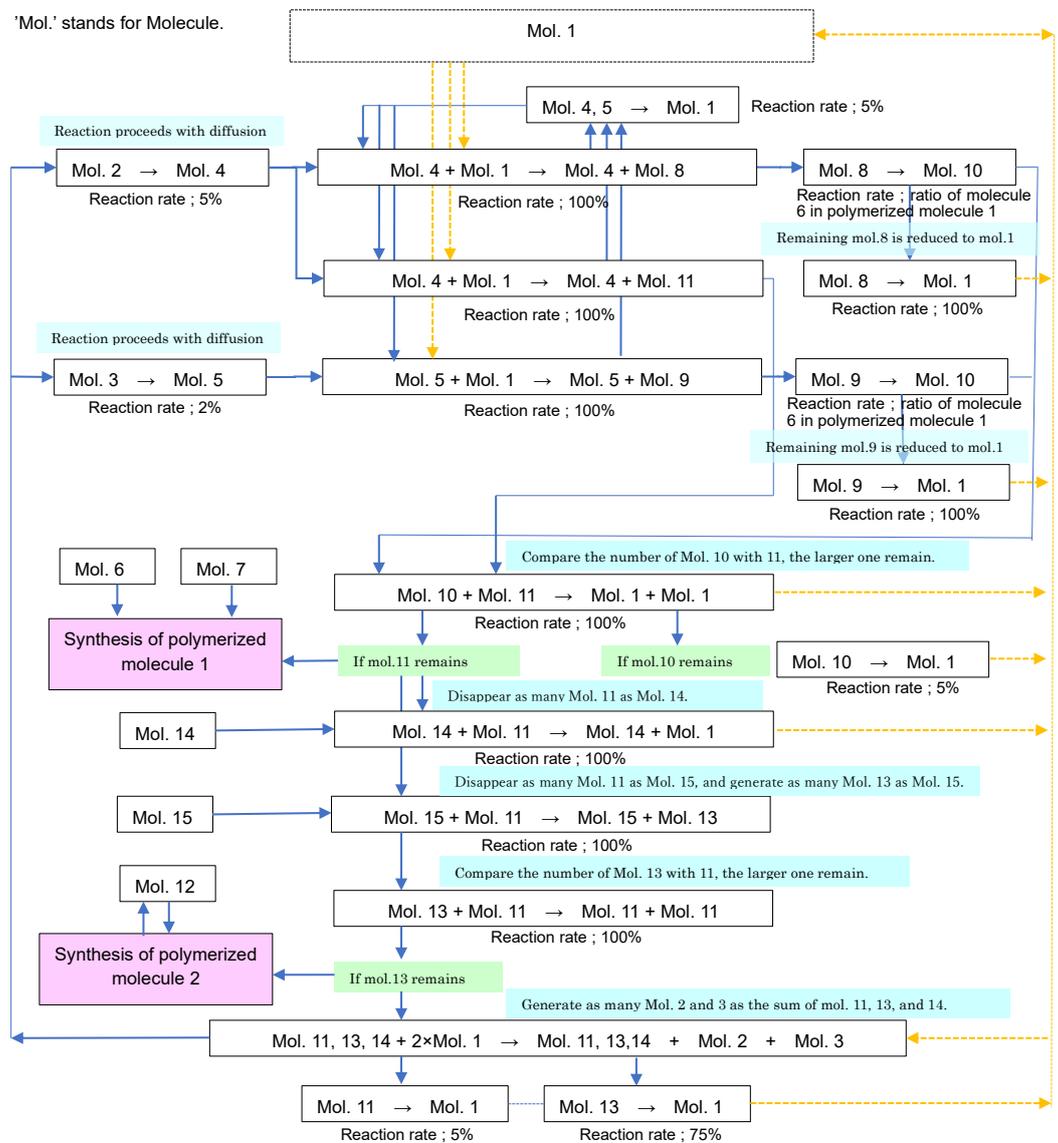

**Figure 4.** Chemical reactions network map. Network of chemical reactions in the developed model. Molecular diffusion and a cyclic network of molecular reactions were constructed to create an algorithm identical to that in the Ishida model [34]. These reaction networks allow for the formation of cell-like shapes and the emergence of self-replication.

*2.3. Method of Evaluation by Entropy*

### 2.3.1. Evaluation Equation for Morphological Entropy

The entropy of the molecular configuration can be calculated from the number of cases, $W$, of molecular configuration using the formula $S = k\, logW$. The number of cases of $m$ molecules to be distributed equally among $X$ spatial cells is as follows:

$$m! / \{(m/X)!\}\, X.$$

Applying Stirling's formula, $log\, n! \sim n\, log\, n - n$:

$$\textit{Maximum configuration entropy} = (m\, log\, m - m) - \{(m/X)log(m/X) - (m/X)\}X,$$

which gives the maximum value of morphological entropy. Using this maximum configuration entropy as a reference, the configuration entropy of the state in which molecules are assembled at a specific location in the spatial field was calculated and the morphological entropy was evaluated using the difference between the two values:

$$\textit{Configuration entropy} = (m\, log\, m - m) - \sum X\, \{(UX)log(UX) - (UX)\}$$

Here, $UX$ is the number of molecules in each cell. Finally, morphological entropy is calculated as follows:

$$\textit{Morphological entropy} = \textit{Maximum configuration entropy} - \textit{Configuration entropy}$$

### 2.3.2. Evaluation Formula for Entropy Production

The amount of entropy production will be the sum of entropy produced by diffusion and chemical reactions, the formulation of which is shown below.

The general formulation of entropy production by diffusion is as follows:

$$S_D = \int_V \sigma_D \mathrm{d}V = \int_V k \frac{D}{C}(\nabla C)^2\, dV,$$

where $D$ is the diffusion coefficient, $C$ is the concentration, and $k$ is Boltzmann's constant. Because the cell space is a virtual field, $k = 1$ and $D = 1$ were assumed. Given that $\nabla C$ is the gradient of concentration, it corresponds to the difference in the number of molecules between neighboring cells. From this, the entropy production in the cell space can be expressed using the following equation:

$$S_D = \sum x\, (\textit{difference in number of molecules between adjacent cells})^2 / \textit{number of molecules}$$

The formulation of a typical entropy production from a chemical reaction is as follows:

$$S_R = \int_V \sigma_R \mathrm{d}V = \int_V k\, ln\frac{v_+}{v_-}(v_+ - v_-)dV,$$

where $v_+$ represents the rate of the forward reaction, $v_-$ is the rate of the reverse reaction, and entropy production follows the rate of reaction. The change in entropy with chemical reactions varies in a complex manner depending on the type of reaction; however, because this model uses an artificial reaction model, the reaction from molecule 1 to other molecular species is assumed to increase entropy production by one unit for each reaction. Then, the reaction to molecule 1 from another molecule was assumed to be a reduction in entropy production by one unit per reaction. The total amount of entropy production of the spatial lattice field was calculated by summing the values of each entropy production.

$$S_R = \sum x(\text{number of reactions that do not return to molecule } 1 - \text{number of reactions from each molecule back to molecule } 1)$$

*2.4. Model Implementation*

The model was programmed in JavaScript language. The program consists of a loop with an initialization part, including a variables setting, a part for calculating chemical reactions and diffusion, and a part for drawing. 2.4.1. Configuration of Lattice Cell Grids

The present model used 2D hexagonal grids wherein the transition rules were simple to apply. Square grids are commonly used in 2D CAs; however, a hexagonal grid is isotropic, whereas a square grid is not isotropic. Because the model includes the process of distributing artificial molecules to adjacent cells, it is simpler to apply when the distances between adjacent cells are equal. Moreover, the model can be applied to a square grid, but the pattern that is created is not isotropic. Models were implemented under the following conditions:

- Calculation field: 100 × 100 cells in hexagonal grids
- Periodic boundary condition

2.4.2. Conditions of Calculations

Molecular diffusion can be controlled by the residual rate of molecules, and the following values were set as the standard cases in the present model. When the residual rate was high, the percentage of molecules remaining in the lattice cell was correspondingly high. However, when the residual rate was low, diffusion occurred more quickly.

(Residual rate of standard case)
Molecule 1 = 0.0
Molecule 2 = 0.75
Molecule 3 = 0.05
Molecule 4 = Equivalent to Molecule 2
Molecule 5 = Equivalent to Molecule 3
Molecule 6 = 0.0
Molecule 7 = 0.0
Molecule 8 = 1.0
Molecule 9 = 1.0
Molecule 10 = 1.0
Molecule 11 = 1.0
Molecule 12 = 0.0
Molecule 13 = 1.0
Molecule 14 = 1.0
Molecule 15 = 1.0
Polymerized molecule 1 = 0.1
Polymerized molecule 2 = 0.75

2.4.3. Initial Conditions

As an initial condition, molecule 1 was placed throughout the lattice with 1,000,000 molecules in each cell, molecule 6 and 7 were present with 50,000 molecules, respectively, and molecule 12 was present with 100,000 molecules. In addition, 100 molecules of molecule 2 and 3 were placed in some cells in the center area of the lattice field (Figure 5A). In addition, 10 polymerized molecule 1, which was polymerized with the ratio of molecules 6 and 7 and shape parameter $w$, were placed in the central region of the space, as shown in Figure 5B.

To determine the standard case value of each parameter, the parameters were adjusted through several rounds of trial and error while referring to the values of Ishida's model [34]. Table 1 shows the values of each parameter for the standard case and the settings for the calculation case with modified parameters for the standard case.

**Table 1.** Standard parameters set and settings for the calculation cases with modified parameters for the standard case

| Focused parameter | Standard set of parameters | Modified parameters for the standard case | | | | | | |
|---|---|---|---|---|---|---|---|---|
| Parameter w | 0.650 | 0.575 | 0.600 | 0.625 | 0.650 | 0.675 | 0.700 | 0.725 |
| Initial arranged number of mol 14 | 3 | | 1 | 2 | 3 | 4 | 5 | |
| Initial arranged number of mol 15 | 9 | | 7 | 8 | 9 | 10 | 11 | |
| Initial configuration of mol.2 and mol.3 | Case1 | | | | Case1 | Case2 | Case3 | |
| Residual rate of mol. 2 | 0.75 | | 0.65 | 0.70 | 0.75 | 0.80 | 0.85 | |
| Residual rate of mol. 3 | 0.05 | | | 0.01 | 0.05 | 0.10 | 0.20 | |
| Reaction rate of mol. 2→ mol.4 | 5 | | | 3 | 5 | 7 | 15 | |
| Reaction rate of mol. 3→ mol.5 | 2 | | | 1 | 2 | 3 | 5 | |
| Residual rate of polymerized molecule 1 | 0.1 | | | 0.05 | 0.1 | 0.15 | 0.2 | |
| Residual rate of polymerized molecule 2 | 0.75 | | | 0.65 | 0.75 | 0.85 | | |
| Removal rate of mol. 4 | 5 | | 0 | 3 | 5 | 7 | | |
| Removal rate of mol. 5 | 5 | | 0 | 3 | 5 | 7 | | |
| Removal rate of mol. 10 | 5 | | 0 | 3 | 5 | 7 | 9 | |
| Removal rate of mol. 11 | 5 | | 0 | 3 | 5 | 7 | 9 | |
| Removal rate of mol. 13 | 75 | | 55 | 65 | 75 | 80 | 95 | |

## 3. Results

*3.1. Emergence of cell-like shapes and their replication patterns*

In this study, 15 types of artificial molecules (molecules 1 to 15) and two types of polymerized molecules (described below) were assumed, as shown in the Materials and Methods section.

- Polymerized molecule 1: molecules 6 and 7 polymerized in the ratio of morphology parameter *w*, which controls the shape of the Turing pattern (spots, stripes, etc.) that appears in equation (1) in the Materials and Methods section. Embedding this information in the lattice space using polymerized molecules is possible. Then, the replication of polymerized molecule 1 represents the propagation of genetic information.

- Polymerized molecule 2: polymerized molecule 2 forms the cell shape (equivalent to membrane molecules).

The calculation results of the constructed model are shown with the distribution of polymerized molecules 1 and 2. In particular, polymerized molecule 2 indicates the cellular region, and the distribution is explained below.

Time series results of polymerized molecule 2 distribution are shown in Figure 5C; the distribution was calculated using the initial conditions shown in Figure 5A and 5B, and the standard parameters shown in Table 1. The reddish color in the figures indicates the distribution of polymerized molecules 2. In the initial stage of the time series, polymerized molecule 2 accumulated in a region in the center of the lattice field, after which this region of molecules was continuously broken up. Simultaneously, $w$ value information was retained in polymerized molecule 1 and transmitted to the field.

Even if the initial conditions and parameter values are the same, the cell-like patterns change with each calculation because the excess molecules that cannot be divided into the six orientations in the molecular diffusion process are assigned to the orientations by random numbers, and parts exist in which the presence or absence of molecular conversion is determined probabilistically based on reaction probabilities. Figure 5D shows five cases of calculations with standard state initial conditions and parameters. In each calculation, the cell replication patterns differ in their details but are similar in that polymerized molecule 2 is replicated sequentially, indicating that the morphogenesis trends are the same.

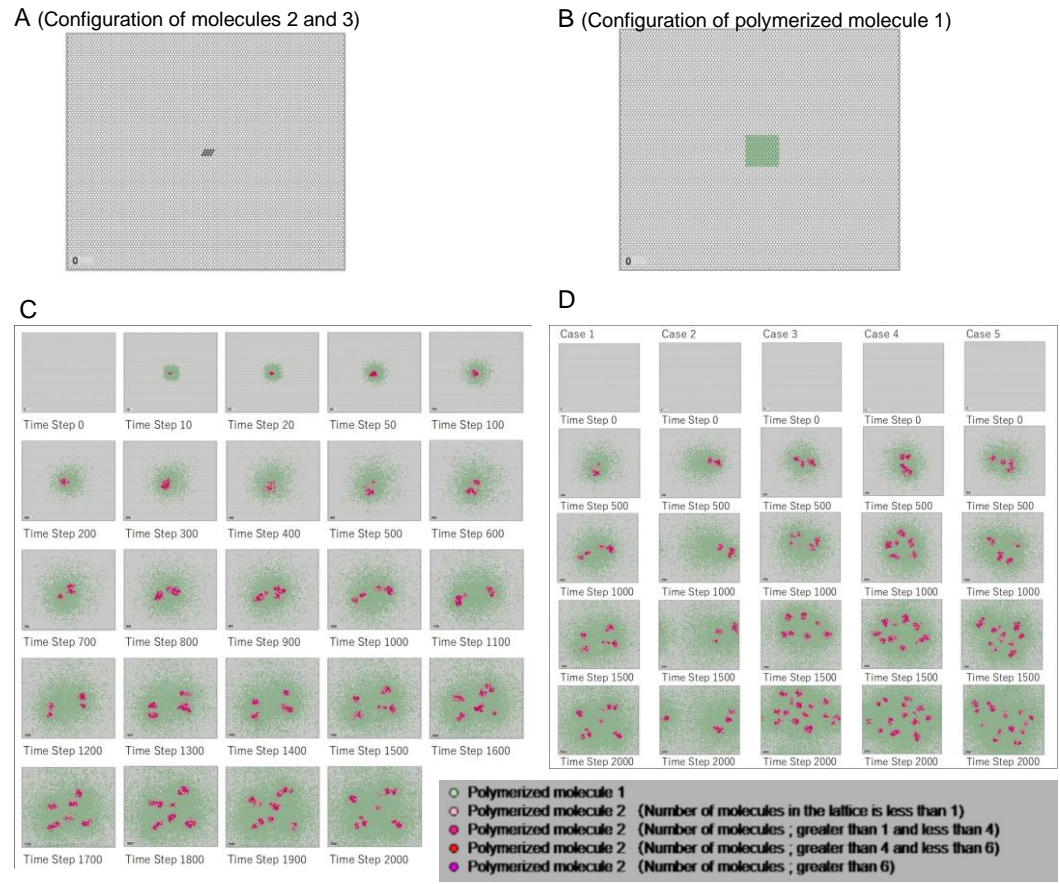

**Figure 5.** Initial configuration of molecules and polymerized molecules, and time series results of emergence patterns and five examples of calculations with standard parameters. **(A)** One hundred molecules of molecule 2 and 3 were placed in some cells in the center area of the lattice field. **(B)** Ten polymerized molecule 1 molecules, which was polymerized with a ratio of molecules 6 and 7 and the shape parameter $w$, were placed in the central region of the lattice space. **(C)** Time series results of the distribution of polymerized molecules 1 and 2 with standard parameters. The reddish color in the figure indicates the distribution of polymerized molecule 2. **(D)** Because random numbers are used in the diffusion and reaction processes, the detailed patterns change with calculations; however, the trend in pattern formation tends to be the same.

The morphology parameter $w$ in the Ishida model [34] is set by the ratio of molecules 6 and 7 in each cell. This parameter controls the shape of the Turing pattern [spots, stripes, etc.; equation (1)]. This ratio is recorded in polymerized molecule 1 and diffused into space. The calculation results obtained with morphological parameter $w$ are shown in Figure 6A.

When the value of $w$ is small, the area of polymerized molecule 2, i.e., the cellular region, spreads out in a mesh-like pattern over the entire space. As the value of $w$ increases, the replication of polymerized molecule 2 becomes intermittent, and at very large values of $w$, the region of polymerized molecules disappears. The same result with $w$ value was shown by Ishida [34], indicating that the cell-like shape can be determined from the $w$ value. The Turing pattern can be formed even with the current multiset chemical model, and cell shape replication can also be achieved by changing the conditions of chemical reactions and polymerization.

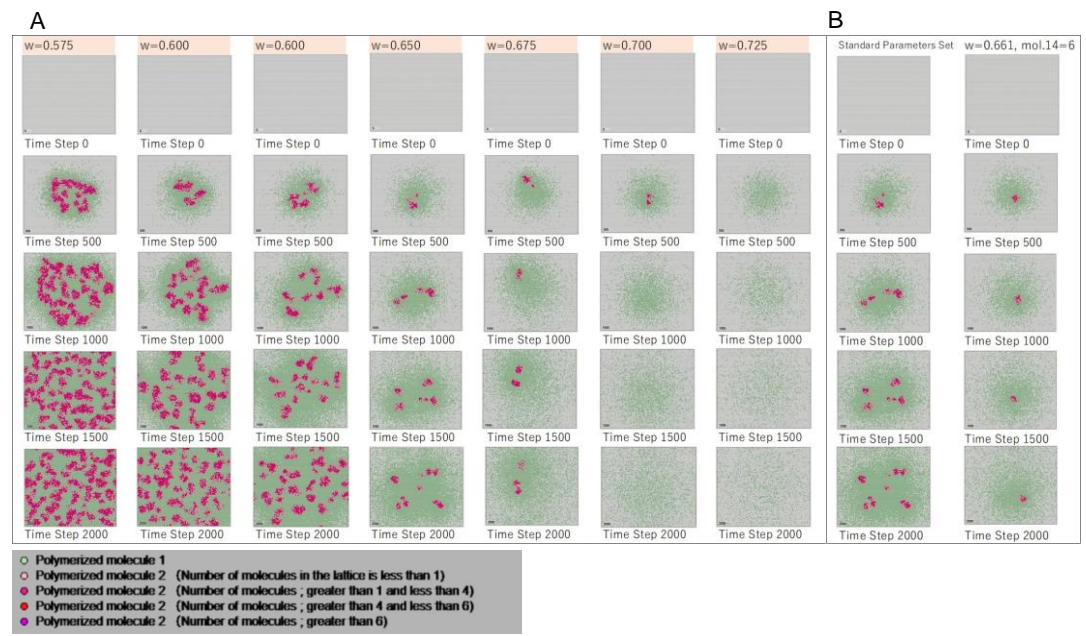

**Figure 6.** Results of morphological patterns with morphological parameter $w$. **(A)** The standard case is $w = 0.650$. When $w$ is small, the area of polymerized molecule 2, which is the cellular region, spreads out in a mesh-like pattern over the entire space. When $w$ increases, the replication of polymerized molecule 2 becomes intermittent. At even larger $w$ values, the region of polymerized molecules disappears. **(B)** Left column: results of the standard case at $w = 0.650$. Right column: results in which the single cell shape is maintained (standard parameters set: $w = 0.661$; number of molecule 14 = 6).

Figure 7A shows the results when the initial configuration of molecules 2 and 3 differed. In Figure 7B, three cases of initial configurations of molecules 2 and 3 are shown. Considering time steps 50–200 in the initial stage of the calculation, no differences were caused by the initial configurations. Thus, a slight change in the initial placement does not lead to a difference in the results.

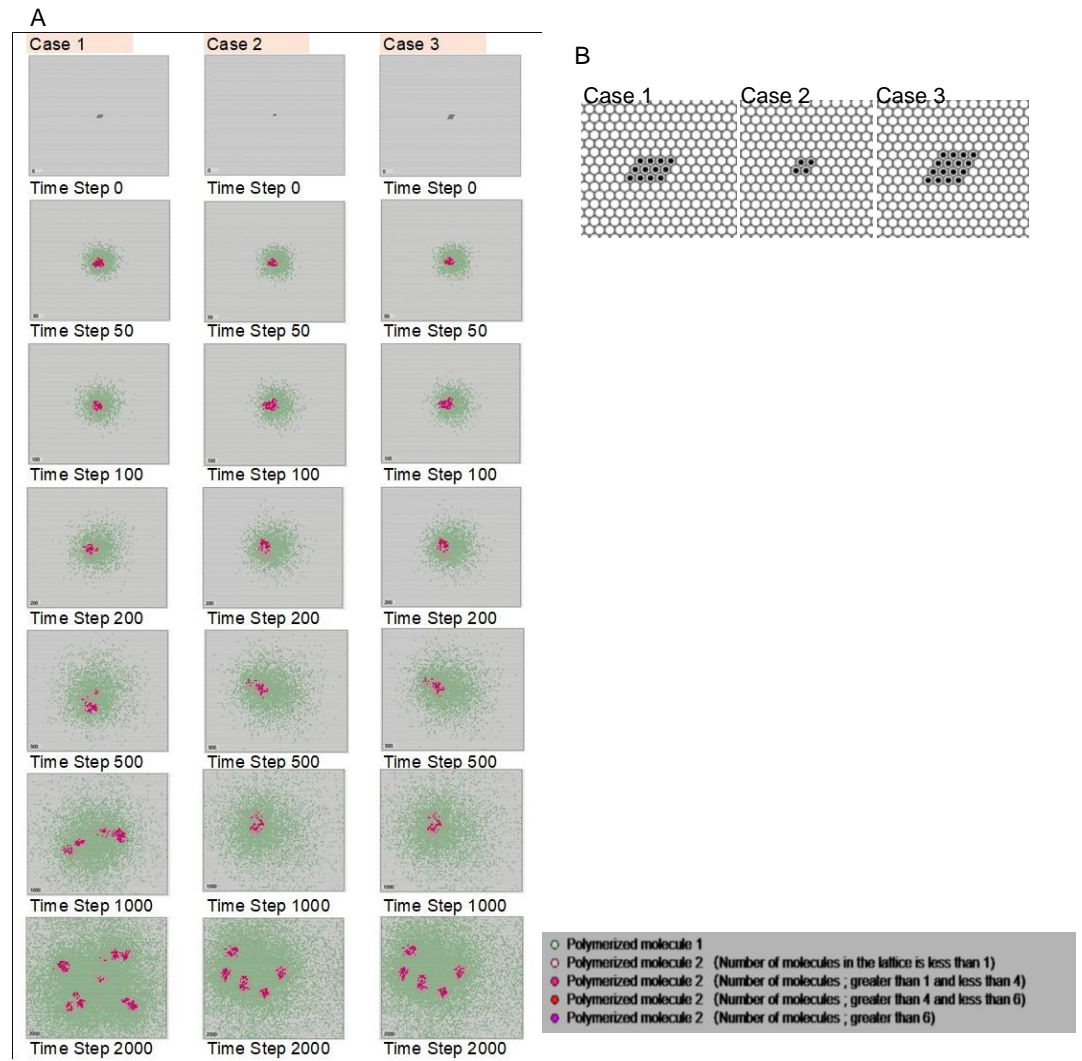

**Figure 7.** calculation results of cases with different initial configurations of molecules 2 and 3. **(A)** Time steps ~50–200, in the initial stage of the calculation, show that no differences occur due to the initial values. **(B)** Three cases of the initial configuration of molecules 2 and 3.

Although many possible calculation cases exist when considering all combinations of parameters in the model, the following cases were those for which a significant change from the standard parameters was detected.

Figure 8A and 8B show the morphological patterns according to the number of molecule 14 and molecule 15, respectively, in the initial arrangement. As described in the Materials and Methods section, after molecules 14 and 15 react with molecule 13, polymerized molecule 2 is controlled and forms in the presence of molecule 13. The calculation results show that the rate of replication varies depending on the number of molecule 14 and molecule 15. The number of these molecules was found to demonstrate a similar effect to that of the $w$ value. Figure 8C shows the results for different residual rates of molecule 2. If these rates were small, molecule 2 diffuses quickly and does not form a clear diffusion difference with molecule 3; consequently, the Turing pattern is not formed. Figure 8D shows the results for different residual rates of molecule 3. In this case, in contrast to molecule 2, as the residual rate of molecule 3 increased, diffusion became slower.

The effects of other parameters on the results are shown in Appendix A (Figure S1 – Figure S9).

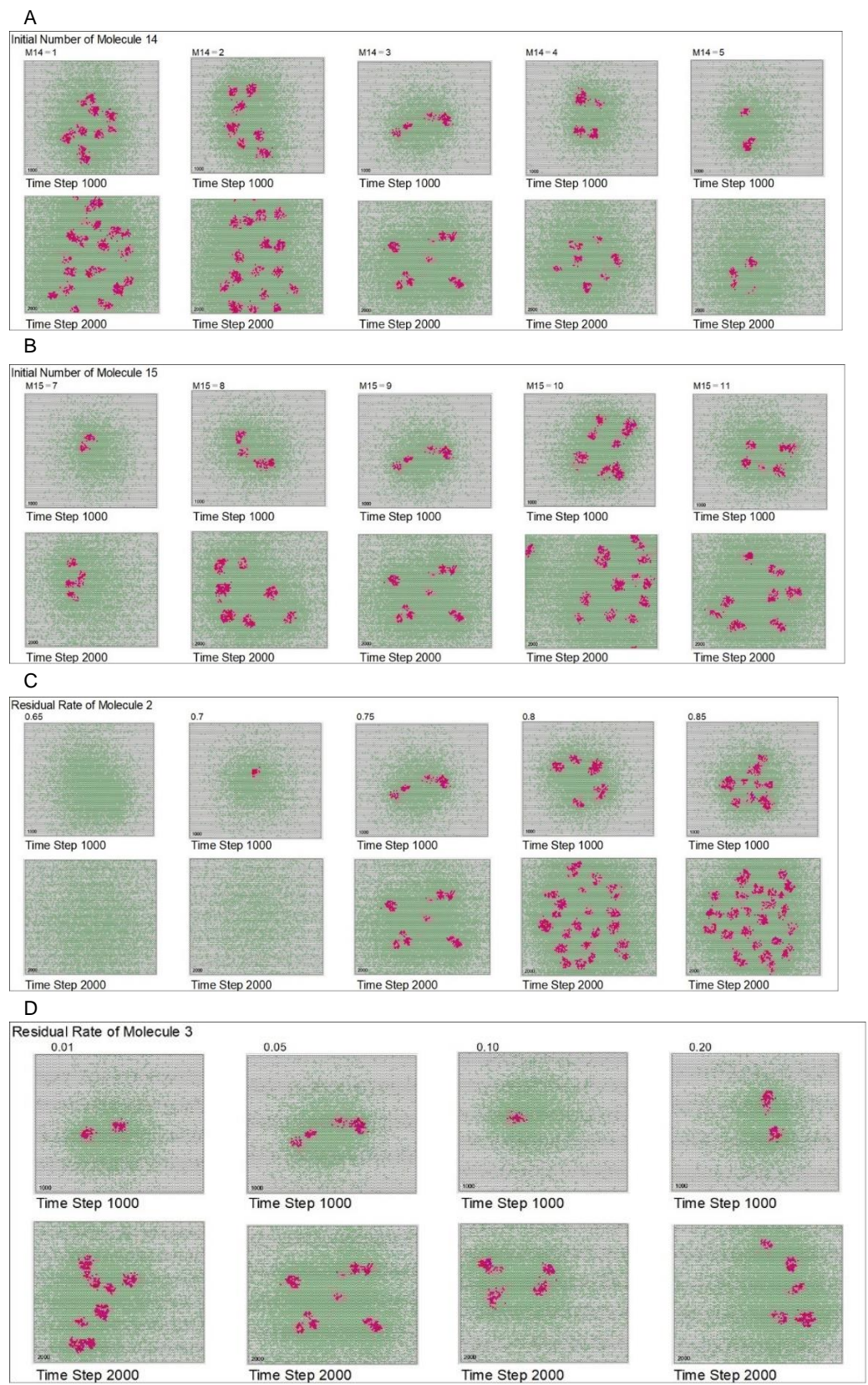

**Figure 8.** Morphological patterns with the change of parameters. **(A)** Morphological patterns

change with the number of molecule 14 in the initial conditions. Each image is the result of time step 1,000 and 2,000 for each condition. Results show that the rate of replication varies depending on the initial number of molecule 14. Additionally, the number of molecule 14 demonstrates a similar effect as that of the $w$ value. **(B)** Morphological patterns change with the number of molecule 15 in the initial conditions. Each image is the result of time step 1,000 and 2,000 for each condition. Results show that the rate of replication varies depending on the initial number of molecule 15. Further, the number of molecule 15 demonstrates a similar effect as that of the $w$ value. **(C)** Morphological patterns change with the residual rates of molecule 2. Each image is the result of time step 1,000 and 2,000 for each condition. When the residual rate of molecule 2 is small, molecule 2 diffuses quickly and does not form a clear diffusion difference with molecule 3. **(D)** Morphological patterns change with the residual rates of molecule 3. Each image is the result of time step 1,000 and 2,000 for each condition. When the residual rate of molecule 3 increases, diffusion becomes slower.

### 3.2. Evaluation by Entropy

If the molecular networks set up described in the Materials and Methods section leads to the emergence of the three conditions of life, the formation of boundaries and self-replication of the cell shape can be visually and qualitatively confirmed by the distribution of polymerized molecule 2. In addition, quantitative evaluation should be possible using entropy as an indicator. Observing life from an entropic perspective, the entropy of form (i.e., the entropy of molecular arrangement, hereafter referred to as "morphological entropy") is kept at a lower state than that in which molecules are uniformly distributed in space (i.e., the state of maximum entropy) because a specific population of molecules can form in a specific region. However, to maintain the form of life and keep entropy low, a constant inflow of high-quality energy and a constant release of low-quality heat energy is necessary. This increases the total entropy, including the external space. Since the developed model is a virtual chemical reaction world, the increase in entropy can be evaluated in terms of entropy production based on chemical reactions.

Figure 9A shows the time series of total entropy generation in which entropy is continuously produced. In the entire lattice space, entropy increases due to continuous molecular reactions. Figure 9B shows the time series variation of total morphological entropy which shows the difference from the maximum entropy state for the placement entropy at each time step, with higher values indicating lower spatial placement entropy. In the process of replication of polymerized molecule 2, morphological entropy decreases; moreover, as the spatial order increases, the morphological entropy value increases.

According to Figure 9A and 9B, the entropy of molecular configuration decreases, whereas the entropy production increases. In addition, as the entropy of the entire lattice space increases, the configuration entropy of the polymerized molecules decreases. This is thought to reproduce the entropic conditions of a situation in which life is active. It is not merely the formation of polymerized molecular forms, but rather a chain of cyclic reactions of the molecules, which is considered to be the expression of metabolism. Figure 9C shows the time series variation of morphological entropy with parameter $w$. When $w$ = 0.60, the pattern structure is spread over the entire lattice space and the value of morphological entropy is high.

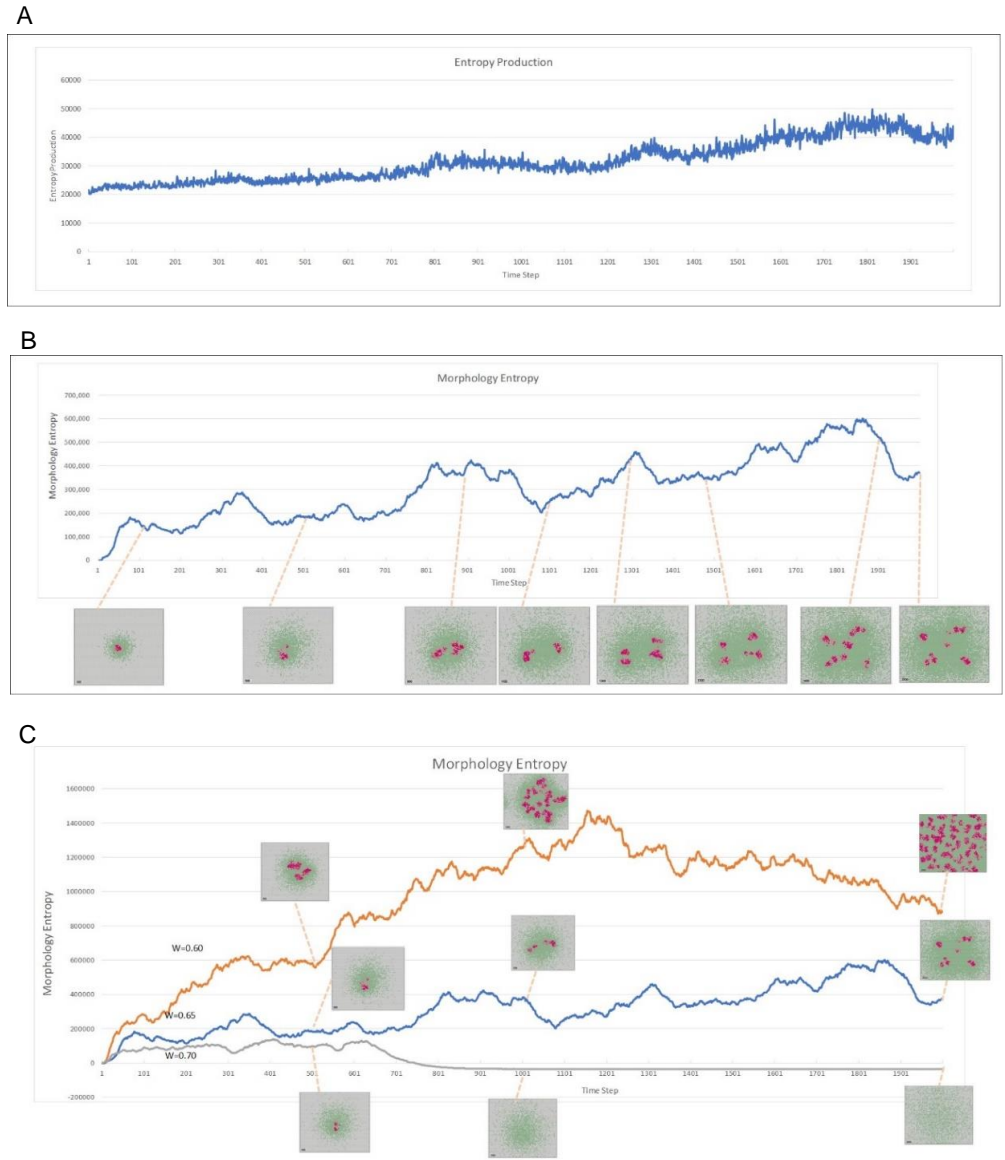

**Figure 9.** Evaluation by entropy. **(A)** Time series of entropy production at standard parameters. Entropy is continuously produced, and entropy increases in the entire lattice space due to continuous molecular reactions. **(B)** Time series variation and pattern of morphological entropy at standard parameters. The difference from the maximum entropy state for the placement entropy is shown at each time step, with higher values in the graph indicating lower spatial placement entropy. In the process of replication of polymerized molecule 2, shape entropy decreases. When the spatial order is high, the entropy value is correspondingly high. **(C)** Time series variation and pattern of morphological entropy with parameter $w$. When $w$ = 0.60, the pattern structure is spread over the entire lattice space, and the value of morphological entropy is high.

## 4. Discussion

The model presented here shows that a self-replicating cell-like region (aggregation of polymerized molecule 2) can be generated by a local chemical reaction network model of virtual molecules while retaining the information of morphological parameter $w$ in the form of the molecular ratio of polymerized molecule 1 in a lattice field. This model does not require information on distant lattice cells, as is the case in the conventional model, and no externally set values exist, such as that of morphological parameter $w$. The only values needed for the calculation are the number of molecules in the field and the values specific to the molecules (e.g., reaction rate and residual rate). Specifically, the ratio of molecules 6 and 7 and the number of molecules 14 and 15 are required.

In addition, these values were found to control the emergent morphological patterns. Furthermore, small changes in the number of molecules 14 and 15 present controlled the degree of division of the cell-like region. These findings may correspond to the functions of these molecules as morphogens or hormones in living organisms, and inducing or suppressing replication may be possible by locally adjusting the number of such molecules. For example, by adjusting and varying the number of molecule 14 and the $w$ value, creating the stational region of polymerized molecule 2 in the center of the lattice space or inducing replication is possible. Figure 6B shows an example case, comparing the results of the standard case with the results of the calculation when the number of molecule 14 = 6 and $w$ = 0.661.

The same result can also be achieved by changing the ratio of molecules 6 and 7 in polymerized molecule 1. In the model, polymerized molecule 1 was set to replicate, whereas the ratio of molecules 6 and 7 in this molecule under initial conditions was maintained. Indeed, the same ratio was maintained throughout the space, but it would also be possible to represent evolution by allowing this ratio to change through mutation or other means. For example, if an environment is assumed in which the number of molecule 14 and 15 locally fluctuates, by mutating the ratio of molecules 6 and 7 in polymerized molecule 1 during the process of division, the molecule rate that survives best in such an environment may emerge. Testing such a change in ratio will be the subject of future research.

In contrast to conventional CA models, such as the Ishida model [34], which are set up with meta-rule and numerical values and only show the formation of abstract state patterns in lattice cell space, the model presented here is essentially different because it considers the formation and degradation of polymerized molecules.

Evaluation by entropy values also showed that the morphology entropy of the molecular arrangement remains low, despite entropy constantly increasing throughout the lattice space. Thus, to maintain a low morphology entropy, an increase in entropy production occurs throughout the system, which reproduces a structure equivalent to that of the entropy of living organisms. Therefore, the region in which polymerized molecule 2 accumulates in the lattice space may be considered "alive."

These results demonstrate that simple networks of chemical reactions can potentially produce a phenomenon including the basic requirements of life: the formation of domains, replication, metabolism, and the inheritance of morphological information. In the various conventional hypotheses for the emergence of life, even if the materials necessary for life were available or simple membranes were formed, how these materials could simultaneously acquire the conditions for life was unclear. However, the current model shows how these conditions can be acquired together and provides insights into the process in which such conditions were established in the first life forms.

Furthermore, the model described here is constructed solely from molecular reactions and molecular polymerization and decomposition; if chemical reactions under similar conditions could be realized *in vitro*, the possibility exists that phenomena similar to that shown in this model could be produced. Therefore, the model may provide a pathway to develop a methodology for experimentally creating life.

Using the current model, the scenario of life emergence could be further clarified, and conventional hypotheses could be tested. For example, a specific scenario could be used in the model to examine the processes by which the first RNA formed (which is a challenge in the RNA world hypothesis) and RNA acquired metabolism and membranes. In addition, the model was able to show the formation of regions in which polymerized molecules are assembled. If similar accumulation of polymerized molecules can be demonstrated experimentally, creating phase-separated regions will be possible. In recent years, various phase-separated regions have been discovered within cells [37]. A phase-separated region of polymerized molecules without a membrane may have been the starting point of the first life. If a phase-separated region of macromolecules is formed, an environment favorable to dehydration reactions could be constructed within the region, and the synthesis of peptides and nucleic acids may advance. If polymerization of peptides and nucleic acids proceeds and catalysts are generated in these regions, a reaction loop that produces the materials for membrane formation might also be created and the formation of membranes with improved reliability could occur. Thus, if the conditions shown in the present model can be realized, the aforementioned region could potentially be maintained by metabolism inside and outside the cellular region and self-replication could be controlled. If the energy inflow to sustain the reaction and the supply of raw material molecules to form the region continues, the cell-like region could then be sustained. This could form the basis for the emergence of complex polymers and other materials.

As for an informer that holds genetic information, the current model also assumes a polymerized molecule with a simple composition ratio comprising two molecule types. Such an informant is a more likely first form than RNA, which holds complex information as code. In a more natural scenario, a self-replicator that carries information regarding the number of molecules or compositional ratio of molecules in polymerized molecules may have arisen first.

When a sequence of such simple polymerized molecules acquires catalytic functionality incidentally, creating an efficient new reaction network in the cellular region is possible. This could lead to the formation of membranes and the advantageous organization of specific molecular sequences. Furthermore, as some of the molecular sequences are processed as information, they could evolve into informants that become primordial RNA. The present results reinforce not only the RNA world hypothesis but also apply to the protein world hypothesis and/or metabolic first hypothesis.

In his book [38], Nick Lane explains the origin of life in terms of the porous compartments inside alkaline hydrothermal vents acting as the first cells in which proton gradients at the acid–alkaline interface generated energy for cellular activity and produced complex materials. However, the process by which cells acquired mechanisms to control replication while retaining genetic informants after leaving the porous compartment remains to be elucidated. The present model could perhaps make the scenario described by Nick Lane more concrete.

In his book [39], Kauffman describes the essence of life as "bound closed circuits," which is a concept first introduced by Maël Montévil and Matteo Mossio [40]. Kauffman states that a nonequilibrium self-constructing system is made possible by a chain of thermodynamic bound reactions that form a closed circuit. However, he only explains abstract concepts and does not present concrete simulations or experiments. In the model presented here, individual chemical reactions bind the next reaction, creating an overall cyclical reaction network. By applying processes equivalent to Turing's model of morphogenesis to these reaction networks, I was able to show the emergence, replication, and metabolism of life forms. Thus, using the model, demonstrating Kaufman's hypothesis of life emergence in a concrete simulation may be possible.

Although the current model is simple, as the number of molecule types increases and the degrees of freedom of the reaction increase, with sufficient energy supply to sustain the reaction, new reaction paths and multiple reaction loops may be formed. This could potentially reproduce the evolution of more complex cells.

Based on the results of this model and considering its feasibility in relation to real chemical reactions, considering the feasibility of cellular emergence in a more realistic manner will also be necessary, e.g., by applying the model to a biochemical reaction model based on MD. In addition, simulations must be conducted in three dimensions.

**5. Conclusions**

In summary, by modeling chemical reactions with 15 molecule types and two types of polymerized molecules and using the morphogenesis rule of the Turing model, it was able to model and demonstrate the process of emergence of a cell-like form with the three conditions except evolution ability. It is believed that the model will help facilitate a discussion on "how the four conditions necessary for life were constructed." Moreover, this model will allow us to revisit and refine each of the existing hypotheses related to the emergence of life.

**Supplementary Materials:** Simulation Video of Standard Parameters case (Fig.2A) at https://youtu.be/DHrNWrYH1Rg. The source code of the simulation model can be downloaded from the following; https://github.com/Takeshi-Ishida/Emergence-simulation-of-biological-cell-like-shapes

**Author Contributions:** The sole author, Takeshi Ishida, conducted all research.

Funding: This research was supported by grants from Japan Society for the Promotion of Science, KAKENHI Grant Number 19K04896.

**Acknowledgments:** The authors would like to thank Enago (www.enago.jp) for the English language review.

**Conflicts of Interest:** The author declares no conflict of interest related to the publication of this paper.

**Appendix A**

Additional results are shown in Appendix A. Figure S1 shows the morphological change with different reaction probabilities from molecule 2 to molecule 4, and Figure S2 shows the morphological change with different reaction probabilities from molecule 3 to molecule 5. It can be seen that the morphological pattern can be changed even by controlling the reaction probability of these molecules.

Figure S3 shows the morphological change with the residual rate of polymerized molecule 1, which is molecule 6 and molecule 7 polymerized in the ratio of shape parameter w. Differences in residual rate affect the distribution of green cells in the figures, but the effect of residuals is not considered significant.

Figure S4 shows the morphological change with the residual rate of polymerized molecule 2, which is molecule 12 polymerized. It seems that the greater the residual rate, the more active the division of the cell-like region.

Figure S5 shows the morphological pattern change with different removal rate of molecule 4, and Figure S6 shows morphological pattern change with different removal rate of molecule 5. In both cases, a removal rate of 0 indicates that the morphological pattern cannot be formed. The fact that the pattern form can also be controlled by the removal rate of tokens is also pointed out in the paper by Ishida [35]. The morphological pattern changes depending on the degree of degradation of the molecules, it suggests that there is some optimal range of removal rates for the formation of replicated patterns.

Figure S7, S8, and S9 show the changes in morphology patterns for different removal rates of molecules 10, 11, and 13, and it can be seen that the morphogenesis changes are not as significant as the results for molecules 4 and 5.

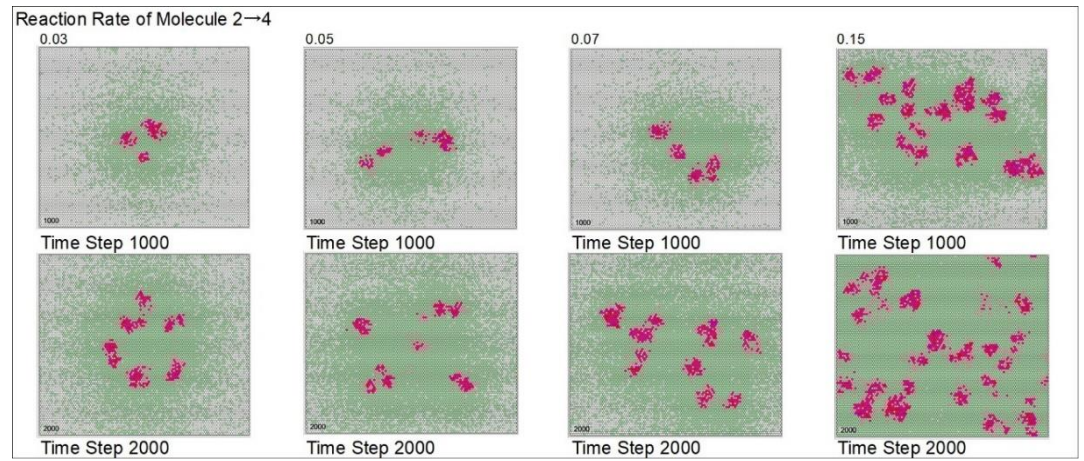

**Figure S1.** Morphological change with different reaction probabilities of reaction from molecule 2 to molecule 4. It can be seen that the morphological pattern can be changed even by controlling the reaction probability of these molecules.

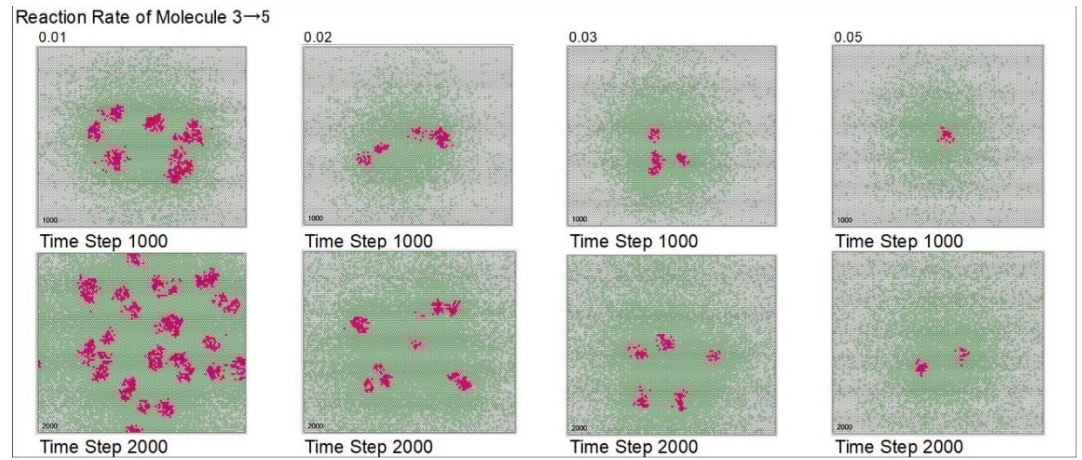

**Figure S2.** Morphological change with different reaction probabilities from molecule 3 to molecule 5. t can be seen that the morphological pattern can be changed even by controlling the reaction probability of these molecules.

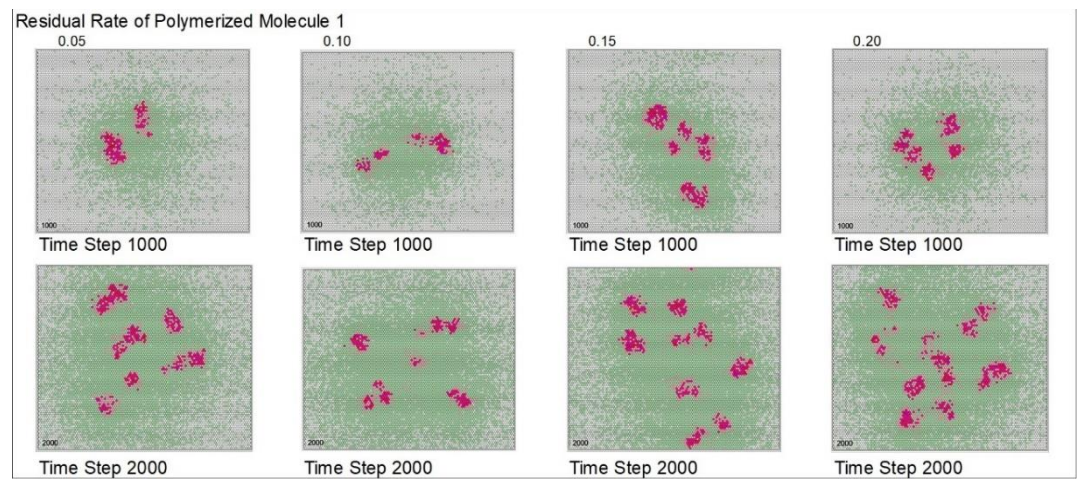

**Figure S3.** Morphological change with the residual rate of polymerized molecule 1, which is polymerized of molecule 6 and molecule 7 in the ratio of shape parameter w. Differences in residual

rate affect the distribution of green cells in the figures, but the effect of residuals change is not considered significant.

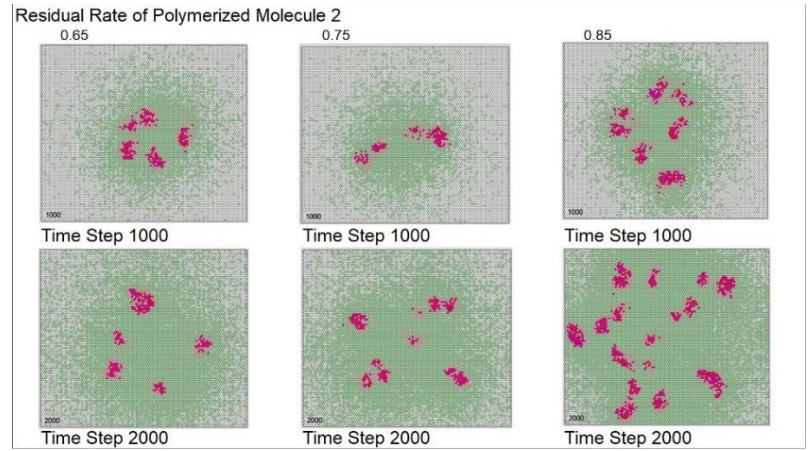

**Figure S4.** Morphological change with the residual rate of polymerized molecule 2, which is polymerized of molecule 12. It seems that the greater the residual rate, the more active the division of the cell-like region.

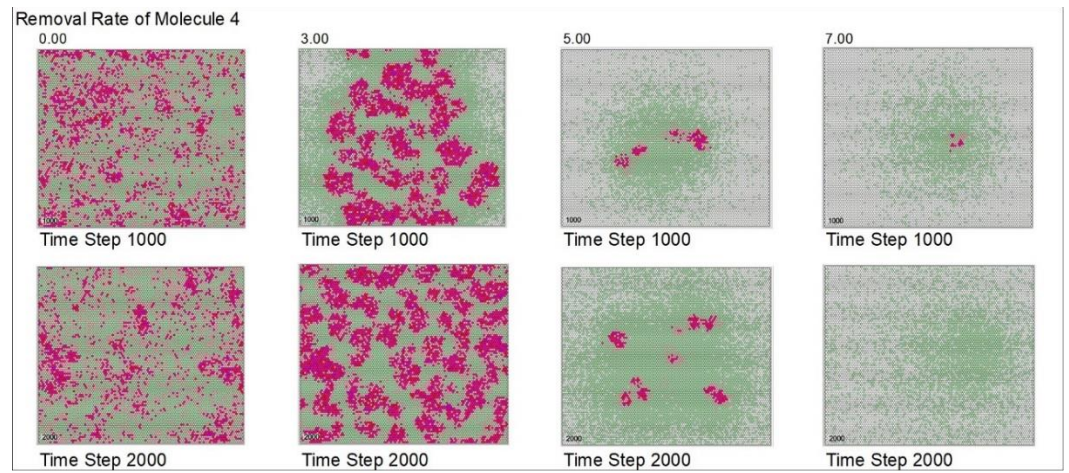

**Figure S5.** Morphological pattern change with different removal rate of molecule 4. A removal rate of 0 indicates that the morphological pattern cannot be formed. The morphological pattern changes depending on the degree of degradation of the molecules.

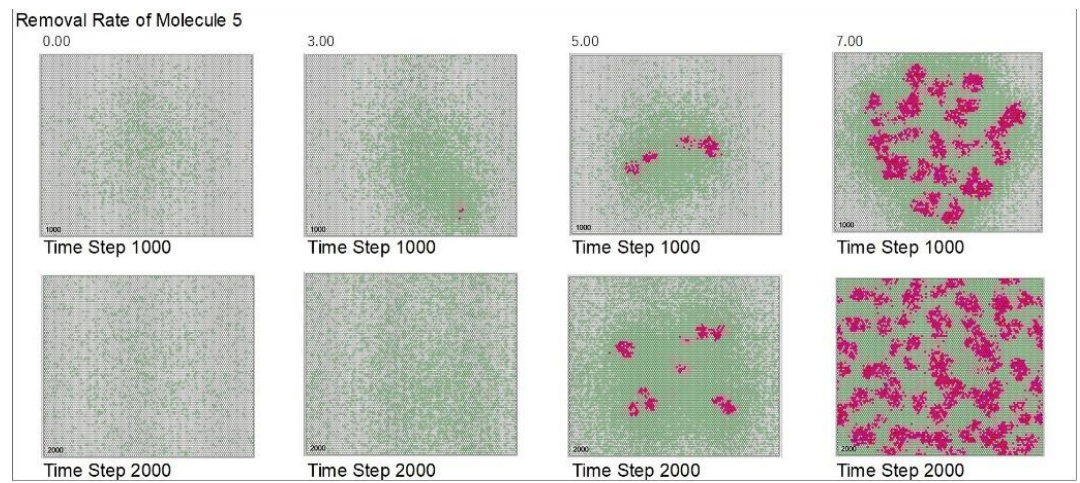

**Figure S6.** Morphological pattern change with different removal rate of molecule 5. A removal rate of 0 indicates that the morphological pattern cannot be formed. The morphological pattern changes depending on the degree of degradation of the molecules.

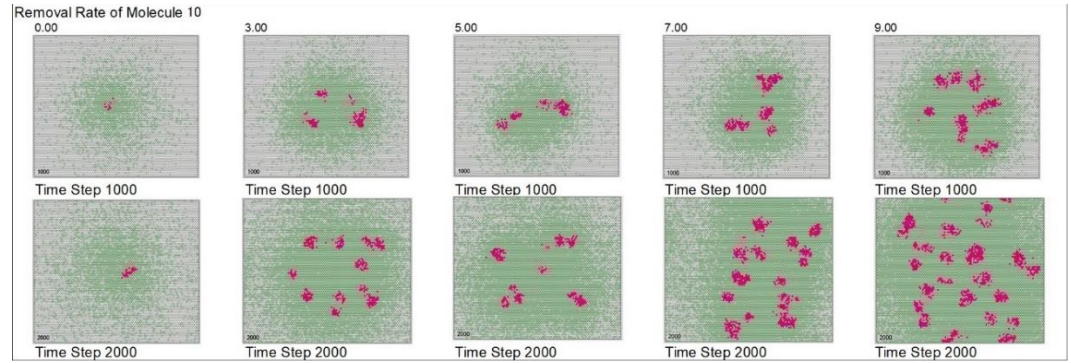

**Figure S7.** Morphological pattern change with different removal rate of molecule 10. I The morphological pattern changes depending on the removal rate of molecule.

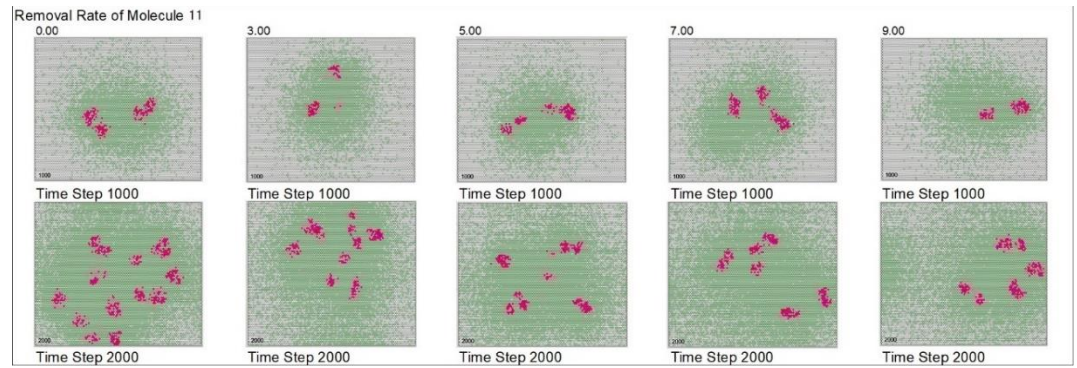

**Figure S8.** Morphological pattern change with different removal rate of molecule 11. It can be seen that the morphogenesis changes are not as significant as the results of Figures S5 and S6 of molecules 4 and 5.

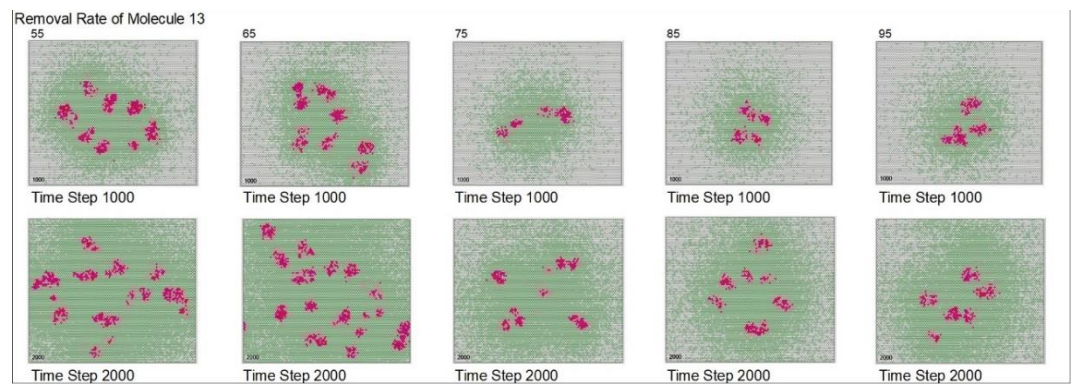

**Figure S9.** Morphological pattern change with different removal rate of molecule 13. It can be seen that the morphogenesis changes are not as significant as the results of Figures S5 and S6 of molecules 4 and 5.